\documentclass[reqno]{amsart}
\usepackage{graphicx,amsmath}
\usepackage{epstopdf}
\usepackage{float}

\def\bq{\mathbf q}
\def\bQ{\mathbf Q}
\def\bJ{\mathbf J}
\def\bb{\mathbf b}
\def\bB{\mathbf B}
\def\re#1{(\ref{#1})}
\def\f{\frac}

\begin{document}

\title{Generalized heat conduction in heat pulse experiments}

\author{Kov\'acs R.$^{13}$ and V\'an P.$^{123}$}

\address{
$^1$Department of Theoretical Physics, Wigner Research Centre for Physics,
Institute for Particle and Nuclear Physics, Budapest, Hungary and 
$^2$Department of Energy Engineering, BME, Budapest, Hungary and 
$^3$Montavid Thermodynamic Research Group}

\date{\today}

\begin{abstract}
A novel equation of heat conduction is derived with the help of a generalized 
entropy current and internal variables. The obtained system of constitutive 
relations is compatible with the momentum series expansion of the kinetic 
theory. The well known Fourier, Maxwell--Cattaneo--Vernotte, Guyer--Krumhansl, 
Jeffreys--type, and  Cahn--Hilliard type equations are derived as special 
cases. 

Some remarkable properties of solutions of the general equation are 
demonstrated with heat pulse initial 
and boundary conditions. A simple numerical method is developed and its 
stability is proved. Apparent faster than Fourier pulse propagation is 
calculated in the over-diffusion regime.
\end{abstract}
\maketitle

\section{Introduction}

 Recently a common generalization of the 
 Fourier, Maxwell-Cattaneo-Vernotte, Guyer-Krumhansl, Jeffreys-type and 
 Green-Naghdi heat conduction equations was derived in the framework of 
 non-equilibrium  thermodynamics \cite{VanFul12a}. Then experimental and 
 theoretical studies  were performed in order to understand the role of 
 different terms and  also the  possibility of detecting non-Fourier effects 
 \cite{CzeEta13p1,CzeEta13p2,VanEta13p1}. According to the basic hypothesis of 
 these investigations, material heterogneities are manifested in additional 
 higher order space and time derivatives in the material functions and result 
 in nonlocal and memory effects (see e.g. 
 \cite{TiaEta13a,SelEta13a,SelEta15a}). However, these 
 effects may be not apparent, the observed phenomena may be Fourier like due 
 to the  universal dissipative nature of the additional terms. Therefore it is 
 important to identify and analyse possible qualitative signatures for 
 experimental observation. 

The non-equilibrium thermodynamical theory of generalized heat conduction of
\cite{VanFul12a} is based on the assumption of a minimal deviation 
from local equilibrium. The deviation is expressed in terms of new 
fields and may appear both in the density and in the current density of the 
entropy:
\begin{itemize}
\item In the entropy a quadratic expression of a vectorial internal variable 
represents the deviation from local equilibrium in the continua 
\cite{Gya77a,Ver97b}. This contribution results in memory effects.
\item A generalization of the entropy current density, with the help of current 
multipliers, represents the deviation of the currents from their local 
equilibrium form \cite{Ver83a,Nyi91a1,Van01a2,CimVan05a}. This contribution 
results in nonlocal effects.
\end{itemize}

The modifications are restricted only by the second law of thermodynamics, do 
not incorporate assumptions about the structure of the continua, 
therefore in this sense the approach is universal \cite{Van13p2,AssEta14a}. 

An important problematic point of the theory is, that the structure of the 
derived system of evolution equations seems to be incompatible with the 
existing theories of Extended Thermodynamics, that is with the hyperbolic 
system of the momentum series expansion hierarchy of the kinetic 
theory \cite{Rug12a}. In particular it does not compatible with the ballistic 
phonons, a well explained non-Fourier propagation mechanism in low temperature 
materials \cite{MulRug98b,JouAta92b}.

In this paper we slightly modify and extend the approach of \cite{VanFul12a} 
introducing the heat flux as a basic field, instead of the general vectorial 
internal variable of \cite{VanFul12a}.  
We also introduce an additional second order tensorial internal variable and 
the corresponding generalization of the entropy flux by current multipliers. 
This way we reproduce the first two levels of the hierarchy of kinetic 
theory in a generalized, phenomenological framework, without any particular
assumptions on the structure of the material (e.g. a rarefied gas). We assume 
only a second law compatible deviation from local equilibrium. 

What we obtain is more general than the corresponding set of equations of 
Extended Thermodynamics, that is the equations obtained from or motivated by 
the hierarchy of moments in kinetic theory. Due to the phenomenological 
assumptions the whole structure is flexible and we can derive several known 
generalizations of the Fourier equation in a uniform framework obtaining 
information regarding their applicability and interrelations. In this respect 
it is remarkable that Green-Naghdi equations 
\cite{GreNag91a,BarSte08a,BarFav14a} are obtained as 
well as Cahn-Hilliard type heat conduction \cite{CahHil58a,ForAme08a}. These 
heat conduction models were justified by rigorous mathematical methods but not 
related to Extended Thermodynamics.

An other important property of our approach is that old paradoxes and 
reservations regarding some forms of heat conduction are shown in a new light. 
For example the well discussed paradox of heat waves with negative values of 
temperature of the Maxwell-Cattaneo-Vernotte and the Jeffreys-type equations 
(see e.g. in \cite{KorBer98a,Ruk14a}) seems to be removed simply because 
thermodynamics requires the gradient of the reciprocal temperature instead of 
the gradient of the temperature in the related terms of the equations. 

This paper focuses on 
the problem of observability of non-Fouirer heat conduction from a theoretical 
point of view. Solving generalized heat conduction models with heat pulse 
initial and boundary conditions will demonstrate that Fourier type solutions 
may appear unexpectedly and therefore  in addition of wavelike effects one may 
look for other observable benchmarks of heat conduction beyond Fourier. 

In the next section we introduce the theory and derive the heat conduction 
equation up to the second current multiplier and show some known 
particular cases. In the third section we introduce a simple finite difference 
numerical method to solve the set of equations. Finally we show some 
demonstrative solutions of the equations on the example of laser flash 
experiment in order to identify possible non-Fourier effects.

\section{Non-equilibrium thermodynamics of heat conduction}

In this paper we restrict ourselves to rigid heat conductors, therefore the 
time derivatives are partial and the density of the material is constant. Our 
starting point is the balance of internal energy:
\begin{equation}
\partial_t e + \nabla\cdot \mathbf q =0.
\label{inten_bal}\end{equation}

Here $e$ is the density of the internal energy, and $\mathbf q$ is the heat 
flux, the current density of the internal energy. $\partial_t$ denotes the 
partial time derivative and $\nabla$ with the central dot is the divergence, 
$\nabla\cdot \mathbf q = tr(\nabla\mathbf q)$.

The second law is given in the following form
\begin{equation}
\partial_t s + \nabla\cdot \mathbf J \geq 0.
\label{s_bal}\end{equation}
Here $s$ is the entropy density and ${\bf J}$ is the entropy current density 
vector. 
For modeling phenomena beyond local equilibrium, we introduce the heat flux 
$\mathbf q$ as basic field variable and also a second order tensorial internal 
variable denoted by $\mathbf Q$. The advantage of using the heat flux as basic 
field quantity instead of a vectorial internal variable of the treatment in 
\cite{VanFul12a} is the easier comparison with Extended Thermodynamics. The  
deviation from local equilibrium will be characterized by two basic 
constitutive hypotheses:
\begin{itemize}
\item We assume a quadratic dependence of the entropy density on the additional 
fields \cite{Gya77a}:
\begin{equation}
s(e, \bq, \bQ) = s_{eq}(e) - \frac{m_1}{2} \bq\cdot \bq- \frac{m_2}{2} \bQ : 
\bQ,
\label{neqs}\end{equation}
where $m_1$ and $m_2$ are positive constant material coefficients.  This is not 
a complete isotropic representation, for the sake of simplicity we have 
introduced a single material coefficient for the second order tensor $\bQ$, 
too. The derivative of the local equilibrium part of the entropy function 
$s_{eq}$ by the internal energy is the reciprocal temperature: $\frac{d 
s_{eq}}{d e} = \frac{1}{T}$ and $\bQ:\bQ = tr(\bQ\cdot \bQ)$. The quadratic 
form may be considered as a first approximation in case of the heat flux and is 
due to the Morse lemma for the internal variable \cite{Ver97b}. The sign is 
determined requiring concave entropy function, that is, 
thermodynamic stability \cite{JouAta92b,MulRug98b}. 
\item We assume that the entropy flux is zero if $\bq=0$ and $\bQ=0$. Therefore 
it can be written in the following form:
\begin{equation}
\bJ = \bb\cdot\bq+\bB :\bQ.
\label{neqJ}\end{equation}
Here $\bb$ is a second order tensorial constitutive function and $\bB$ is a 
third order one. They are the current multipliers introduced by Ny\'iri 
\cite{Nyi91a1}. General aspects of this assumption were 
treated in \cite{Van01a2} and the special case of heat conduction was 
considered in \cite{VanFul12a,SelEta13a}.
\end{itemize}

Now the basic fields are $T,\bq$ and $\bQ$, the constitutive 
functions are $\bb$ and $\bB$. The entropy production is:
\begin{eqnarray}
\partial_t s+ \nabla\cdot \bJ 
&=&	-\frac{1}{T}\nabla\cdot \bq - 
	m_1\bq\cdot\partial_t\bq -
	m_2\bQ :\partial_t\bQ + 
	\bb:\nabla\bq + \nonumber\\
& &	\bq\cdot(\nabla\cdot\bb) +
	\bB\vdots\nabla\bQ + 
	\bQ: (\nabla\cdot\bB) \nonumber\\
&=& \left(\bb - \frac{1}{T} \mathbf I \right): \nabla\bq + 
	\left(\nabla\cdot\bb -  m_1 \partial_t\bq\right)\cdot \bq + \nonumber\\
& &	\left(\nabla\cdot\bB -  m_2 \partial_t\bQ\right): \bQ +
	\bB \vdots \nabla\bQ \geq 0. 
\label{entrpr}\end{eqnarray}
Here $\mathbf I$ is the unit tensor and the triple dot denotes the full 
contraction of third order tensors. In the last row the first and the third 
terms are products of second order tensors, the second term is the product of 
vectors and the last term is of third order ones.
The time derivatives of the state variables $\bq$ and $\bQ$ represent their 
evolution equations, here they are constitutive quantities. 
Therefore one can identify four thermodynamic forces and currents in the above 
expression and assume linear relationship between them in order to obtain the 
solution of the entropy inequality. 
\begin{center}
\begin{tabular}{c|c|c|c|c}
       &Thermal & Extended thermal & Internal & Extended internal \\ \hline
Fluxes & $\nabla\cdot \bb - m_1 \partial_t \bq$ & 
    $\bb -\frac{1}{T}\mathbf I $ & 
    $\nabla\cdot \bB - m_2 \partial_t \bQ$ &
    $\bB $\\ \hline
Forces &$ \bq $ &
    $\nabla \bq$ &
    $\bQ$ &
    $\nabla \bQ$\\
    \end{tabular}\\
\vskip .21cm
{Table 1. Thermodynamic fluxes and forces}\end{center}

The third and fourth force-current pairs are related to the tensorial internal 
variable $\bQ$. In case of isotropic materials only the 
second order tensors can show cross effects (extended thermal and internal 
interactions), the vectorial (thermal) and third order 
tensorial terms (extended internal) are independent. 

In the following we will simplify the treatment and develop the theory in one 
spatial direction. In the one dimensional representation of the tensors we 
remove the boldface letters, and the one 
dimensional spatial derivative is denoted by $\partial_x$.

In this case the entropy production can be rewritten as:
\begin{equation}
\left(b - \frac{1}{T}\right): \partial_x q + 
\left( \partial_x b -  m_1 \partial_t q\right) q + 
\left( \partial_x B -  m_2 \partial_t Q\right)Q + 
B  \partial_x Q \geq 0. 
\label{entrpr1d}\end{equation}

The linear relations between the thermodynamic fluxes and forces 
result in the following constitutive equations:
\begin{eqnarray}
m_1 \partial_t q - \partial_x b &=& -l_1 q, \label{ce1}\\
m_2 \partial_t Q - \partial_x B &=& -k_1 Q + k_{12} \partial_x q, \label{ce2}\\
b- \frac{1}{T} &=& -k_{21} Q + k_{2} \partial_x q, \label{ce3}\\
B &=& n \partial_xQ. \label{ce4}
\end{eqnarray}

The entropy inequality \re{entrpr1d} requires the following inequalities
\begin{equation} 
l_1\geq 0, \quad
k_1\geq 0, \quad
k_2\geq 0, \quad 
n\geq 0, \quad \text{and} \quad
K=k_1k_2-k_{12}k_{21} \geq 0.
\label{coeffreq}\end{equation}

The above set of constitutive equations \re{ce1}-\re{ce4} 
together with the energy balance \re{inten_bal} and the caloric equation of 
state $T(e)$ give a solvable set of equations, with suitable boundary and 
initial conditions. In case of constant coefficients one can easily eliminate 
the current multipliers by substituting them from \re{ce3}-\re{ce4} into 
\re{ce1}-\re{ce2} and obtain:
\begin{eqnarray}
m_1 \partial_t q + l_1 q - k_2 \partial^2_x q &=& 
	\partial_x \f{1}{T}- k_{21}\partial_xQ, \label{ce11}\\
m_2 \partial_t Q +k_1 Q - n\partial^2_x Q &=& k_{12} \partial_x q. \label{ce12}
\end{eqnarray}

Here $\partial_x^i$ denotes the $i$-th partial derivative by $x$. This set of 
equations is similar to the 13 field equations of kinetic 
theory \cite{DreStr93a}.

We can eliminate the internal variable $Q$, too,  and obtain the following 
general constitutive equation for the heat flux $q$:
\begin{gather}
m_1m_2 \partial_{tt} q + (m_2l_1+m_1k_1)\partial_t q - (m_1 n+m_2 
k_2)\partial_{xxt}q
+\nonumber\\ nk_2 \partial_x^4 q - (l_1 n+K)\partial_{xx} q +k_1l_1 q 
=\nonumber\\
m_2\partial_{xt} \frac{1}{T} +k_1 \partial_x \frac{1}{T} - n \partial_x^3 
\frac{1}{T}.
\label{genhc}\end{gather}
In this equation the coefficients are nonnegative according to the second law. 
One can distinguish between the following special cases:

\subsection{} $n=0$. Then the second current multiplier $B$ is eliminated and 
therefore  the highest order spatial derivatives of $q$ and $T$ are missing.
\begin{gather}
m_1m_2 \partial_{tt} q + (m_2l_1+m_1k_1)\partial_t q - m_2 k_2\partial_{xxt}q 
 +k_1l_1 q -K\partial_{xx} q  = \nonumber\\
 m_2\partial_{xt}\frac{1}{T}+k_1 \partial_x \frac{1}{T}.
\label{genhc1}\end{gather}

\subsection{Ballistic-conductive}\label{balcond} $n=k_2=0$. Then we obtain
\begin{gather}
m_1m_2 \partial_{tt} q + 
	(m_2l_1+m_1k_1)\partial_t q +
	k_1l_1 q - 
	k_{12}k_{21} \partial_{xx} q =
 m_2\partial_{xt}\frac{1}{T}+
	k_1 \partial_x \frac{1}{T}.
\label{balhc1}\end{gather}
We will see, that this equation may show ballistic propagation. In this case 
either $k_{12}$ or 
$k_{21}$ is not positive because of \re{coeffreq}.

\subsection{Guyer-Krumhansl} $n=m_2=0$. In this case both the tensorial 
internal variable and the corresponding current multiplier is eliminated:
\begin{gather}
m_1k_1\partial_t q + k_1l_1 q - K\partial_{xx} q  = k_1 \partial_x \frac{1}{T}.
\label{genhc2}\end{gather}

\subsection{Generalized} $n=m_1=0$. Then one obtains the generalized heat 
conduction law of \cite{VanFul12a}:
\begin{gather}
m_2l_1\partial_t q - m_2 k_2\partial_{xxt}q - K\partial_{xx} q +k_1l_1 q =
m_2\partial_{xt} \frac{1}{T}+k_1 \partial_x \frac{1}{T}.
\label{genhc3}\end{gather}

\subsection{Cahn-Hilliard type} $n=m_1=m_2=0$. Then one obtains Fourier 
equation extended by the second spatial derivative of the heat flux:
\begin{gather}
 k_1l_1 q- K\partial_{xx} q  =
k_1 \partial_x \frac{1}{T}.
\label{genhc4}\end{gather}
This equation is a Cahn-Hilliard type one, similar to the hypertemperature 
model of Forest and Amestoy except a sign \cite{ForAme08a}.

\subsection{Jeffreys type} $n=m_1=k_2=0$ and $k_{12}$ or $k_{21}=0$.  Then one 
obtains the Jeffreys type like heat conduction law:
\begin{gather}
m_2l_1\partial_t q  +k_1 l_1 q =
m_2\partial_{xt} \frac{1}{T}+k_1 \partial_x \frac{1}{T}.
\label{genhc5}\end{gather}
Strictly speaking it is different of the Jeffreys type equation, because of the 
reciprocal 
temperature instead of the temperature on the right hand side.  

\subsection{MCV} $n=m_2=k_2=0$ and $k_{12}$ or $k_{21}=0$.  Then we get  the 
Maxwell-Cattaneo-Vernotte equation:
\begin{gather}
m_1\partial_t q  +l_1 q = \partial_x \frac{1}{T}.
\label{genhc6}\end{gather}
\subsection{Fourier} $n=m_2=m_1=k_2=0$ and $k_{12}$ or $k_{21}=0$.  Then 
eliminating almost everything one obtains the thermodynamic form of the Fourier 
law:
\begin{gather}
l_1 q = \partial_x \frac{1}{T} = - \frac{1}{l_1 T^2} \partial_x T.
\label{genhc7}\end{gather}

Here $\lambda = \frac{1}{l_1 T^2}$ is the Fourier heat conduction coefficient.

\section{Heat pulse experiment}

Heat pulse experiments were important in the discovery of second sound and 
ballistic propagation phenomena (see e.g. \cite{Pes44a,JacWal71a}) and in the 
framework of laser flash method it is an important part 
of standard engineering practice \cite{ParEta61a,BabEta11a,CzeEta13a}. In this 
section we analyze some consequences of the previous general equations in case 
of heat pulses in order to identify possible qualitative effects. The system of 
equations to be solved are the energy balance 
\re{inten_bal} and the constitutive evolution equations for the heat flux and 
the internal variable \re{ce11}-\re{ce12} with the initial and boundary 
conditions specific for the heat pulse experiment. We 
substitute the caloric equation of state $e=\rho c T$, into the energy balance, 
where $\rho$ is the density and $c$ is the specific heat. In this numerical 
study we restrict ourselves to the ballistic-conductive model, where $n=0$ and 
$k_2=0$. Introducing a convenient notation  for the constitutive equations 
results in:
\begin{eqnarray}
\rho c \partial_t T + \partial_x q &=& 0, \label{es1}\\
\tau_q \partial_t q + q  &=& 
	-\lambda\partial_x T- \kappa_{21}\partial_xQ, \label{es2}\\
\tau_Q \partial_t Q + Q &=& \kappa_{12} \partial_x q. \label{es3}
\end{eqnarray}

Here $\tau_q= \f{m_1}{l_1}$ and $\tau_Q= \f{m_2}{k_1}$ are relaxation times, 
$\lambda = \f{1}{T^2 l_1}$ is the Fourier heat conduction coefficent and 
$\kappa_{21}=\f{k_{21}}{l_1}$, $\kappa_{12}=\f{k_{12}}{k_1}$ are  material 
parameters. 

The front side boundary condition is a heat pulse of the following form:
\begin{center}
$q_0(t)=q(x=0,t)= \left\{ \begin{array}{cc}
q_{max}\left(1-cos\left(2 \pi \cdot \frac{t}{t_p}\right)\right) & 
\textrm{if } 0<t \leq t_p,\\
0 & \textrm{if } t>t_p.
\end{array} \right.  $
\end{center}

At the backside  of the sample, $x=L$, we consider adiabatic insulation, 
therefore $q(x=L,t)=0$. Initially the fields are homogeneous and the initial 
conditions are $T(x,t=0)=T_0$ and $q(x,t=0)=0$.

After these operations we introduce the dimensionless variables  $\hat t, \hat 
x, \hat T, \hat q, \hat Q$ for the time, space, temperature, heat 
flux and internal variable, respectively. 
\begin{eqnarray}
\hat{t} =\frac{\alpha t}{L^2}, \quad &\text{where}& \quad
	\alpha=\frac{\lambda}{\rho c};  \quad
\hat{x}=\frac{x}{L};\nonumber \\
\hat{T}=\frac{T-T_{0}}{T_{end}-T_{0}}, \quad &\text{where}&\quad
		T_{end}=T_{0}+\frac{\bar{q}_0 t_p}{\rho c L};  \nonumber \\
 \hat{q}=\frac{q}{\bar{q}_0}, \quad &\text{where}&\quad
\bar{q}_0=\frac{1}{t_p}  \int_{0}^{t_p} q_{0}(t)dt;\ \ \
\hat{Q} = \sqrt{-\f{\kappa_{12}}{\kappa_{21}}} \bar{q}_0 Q.
 \label{ndvar}\end{eqnarray}

The dimensionless parameters are
\begin{equation}
\hat{\tau}_\Delta =\frac{\alpha t_p}{L^2}; \ 
\hat{\tau}_q 	  = \frac{\alpha \tau_{q}}{L^2}; \  
\hat{\tau}_Q 	  = \frac{\alpha  \tau_{Q}}{L^2}; \
\hat{\kappa} 	  = \f{-\sqrt{\kappa_{12} \kappa_{21}}}{L}.
\end{equation}
Here $\hat{\kappa}>0$ and the coefficient in $\hat Q$ is positive due to the 
entropy inequality. Finally we get the ballistic-conductive equations in a 
non-dimensional form
\begin{eqnarray}
\hat{\tau}_{\Delta}\partial_{\hat t} \hat T + 
	\partial_{\hat x} \hat q &=& 0 \ ,\nonumber \\
\hat{\tau}_q \partial_{\hat t} \hat q + \hat q + 
    \hat{\tau}_{\Delta}\partial_{\hat x}\hat T + 		
    \hat{\kappa}\partial_{\hat x}\hat Q &=& 0 \ 
    ,\label{nd_balcond}\\
\hat{\tau}_Q \partial_{\hat t}\hat Q +\hat Q + 
	\hat \kappa \partial_{\hat x}\hat q &=& 0 .\nonumber 
\end{eqnarray}
If $\hat{\tau}_Q=0$, then the Guyer-Krumhansl system is obtained.
 
The corresponding initial conditions are $\hat T(\hat x,\hat t=0)=0$ and $\hat 
q(\hat x,\hat t=0)=0$. The boundary conditions are given for the heat flux 
only. At the rear side $\hat q(\hat x = 1,\hat t)=0$ and the heat pulse at the 
front side is

\begin{center}
$\hat q(\hat x=0,\hat t)= \left\{ \begin{array}{cc}
\left(1-\cos \left(2 \pi \frac{\hat t}{\hat \tau_{\Delta}}\right)\right) & 
\textrm{if } 0<\hat t \leq\hat \tau_{\Delta},\\
0 & \textrm{if } \hat t>\hat \tau_{\Delta}. 
\end{array} \right. $
\end{center}

The above set of initial and boundary conditions will be sufficient for the 
numerical solution of the problem.

\section{Numerical method}

In this and also in the  following sections the previously introduced 
dimensionless variables and parameters will be used without hat. Now we analyze 
a finite difference method of solution of the above mathematical problem. 

\begin{figure}[ht]
\centering
\includegraphics[width=10cm,height=7cm]{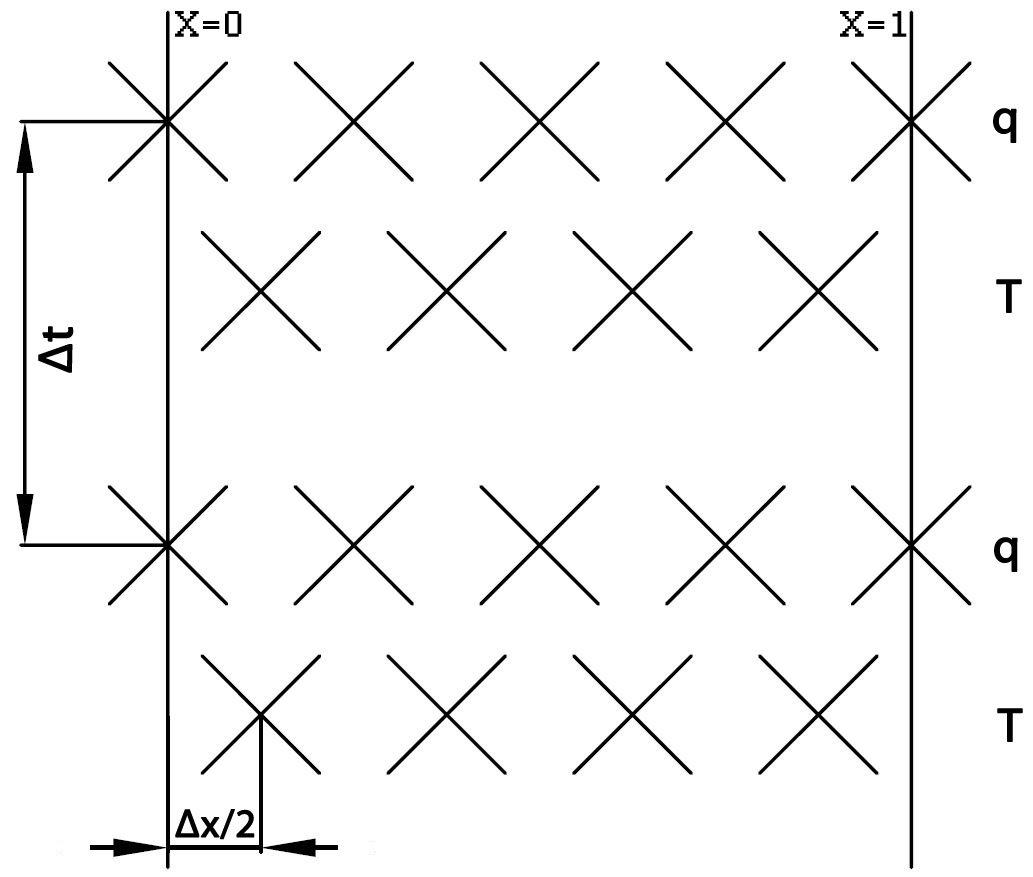}
\caption{Discretization method. The crosses denote the points of space 
discretization.}
\label{fig:disc}
\end{figure}

We apply a finite difference scheme which is forward in time and one-sided in 
space. In fact, it uses the values like a centered scheme in space which holds 
only for shifted fields. There are two kind of discretized fields, one of them 
is covering the space interval from $(0, 1)$, while the others are shifted by 
$\frac{\Delta x}{2}$ (see figure \ref{fig:disc}).

The essential aspect of this method is that we can neglect the boundary 
conditions of shifted fields, because the points of these fields correspond to 
the middle point of each cell. In the particular case of the laser flash 
experiment it is straightforward to use only the heat flux boundary condition. 
It is noted that the fields can be interchanged, the shifting is not necessary 
and it can be mixed, e.g. if one want to define 
conditions both for temperature and heat flux, then only the current 
density of heat flux should be shifted. 

The discrete form of the balance of internal energy reads as:
\begin{equation*}
T^{n+1}_{j}=T^{n} - \frac{\Delta t}{\tau_{\Delta} \Delta 
x}(q^{n}_{j+1}-q^{n}_{j}) .
\end{equation*}
The discrete constitutive equations are:
\begin{eqnarray*}
q^{n+1}_{j} &=& q^{n}_{j} (1 - \frac{\Delta t}{\tau_{q}}) - \frac{\Delta t 
\tau_{\Delta}}{\tau_{q} \Delta x}(T^{n}_{j} - T^{n}_{j-1}) - \frac{\kappa 
\Delta t}{\tau_{q} \Delta x} (Q^{n}_{j}-Q^{n}_{j-1}), \\
Q^{n+1}_{j} &=& Q^{n}_{j} - \frac{ \Delta t}{\tau_{Q}} Q^{n}_{j} + 
\frac{\kappa \Delta t}{\tau_{Q} \Delta x} (q^{n}_{j+1}-q^{n}_{j}).
\end{eqnarray*}
Here $n$ and $j$ are the time and the space indeces.

The scheme is explicit therefore  the stability is an essential side of the 
analysis. We have applied the von Neumann method, i.e. we assumed the solution 
of the difference equation in the form $T^{n}_{j}= \psi^{n} \cdot e^{ikj \Delta 
x}$, where $\psi$ is the growth factor with stability condition $|\psi| \leq 
1$, $i$ is  the imaginary unit, $k$ is the wave number. After substitution we 
get the system of linear algebraic equations:
\begin{equation*}
\bf{M} \cdot 
\left ( \begin{array}{c}
T \\
q \\
Q
\end{array} \right ) =0, 
\end{equation*}
where 
$$\bf{M} = \left ( \begin{array}{ccc} 
1-\psi & -\frac{\Delta t}{\tau_{\Delta} \Delta x}(e^{ik \Delta x} -1) & 0 \\
-\frac{\Delta t \tau_{\Delta}}{\tau_{q} \Delta x}(1-e^{-ik \Delta x}) & 1- 
\frac{\Delta t}{\tau_{q}} - \psi & -\frac{\kappa \Delta t}{\tau_{q} 
\Delta x}(1-e^{-ik \Delta x}) \\
0 & \frac{\kappa \Delta t}{\tau_{Q} \Delta x}(e^{ik \Delta x} -1) & 1- 
\frac{\Delta t}{\tau_{Q}} - \psi 
\end{array} \right ). $$

The characteristic equation of the system can be calculated from $\det \bf{M} 
=$0 and Jury criteria is applied for stability analysis. 

\subsection{Jury criteria}
If the characteristic equation is given in the form 
\begin{equation}
F( \psi )=c_3 \psi^3 + c_2 \psi^2 + c_1 \psi^1 + c_0 
\end{equation}
where $c_i ,(i=0,1,2,3)$, then the related conditions of stability are 
\cite{Jur74b}
\begin{enumerate}
\item $F( \psi =1) >0$,
\item $F( \psi =-1)<0$,
\item $|c_0|<c_3$,
\item $|b_0 |>|b_2|$, where 
$b_0= \left | \begin{array}{cc}
c_0 & c_3 \\
c_3 & c_0
\end{array} \right | $ and
$b_2= \left | \begin{array}{cc}
c_0 & c_1 \\
c_3 & c_2
\end{array} \right | $.
\end{enumerate}
The analytical calculation of these conditions is straightforward, but 
inconvenient even 
in this size, so the numerical evaluation of the criteria is recommended. The 
conditions are checked during the calculations. We did not encounter remarkable 
stability problems with this discretization.

\section{Solutions}

In this section we classify the solutions of the Guyer-Krumhansl equation  as 
Fourier-like, MCV-like (under-diffusive) and GK-like (over-diffusive). We also 
present solutions of the ballistic-conductive 
system. The characteristics of the solution are demonstrated separately on the 
backside temperature profile. 

\subsection{Solutions of the GK-equation}
Guyer-Krumhansl equation is a special form of  \re{nd_balcond} if there are no 
inertial effects in the propagation of the internal variable $Q$, that is 
$\tau_Q = 0$.

\begin{figure}[b]
\centering
\includegraphics[width=12cm, height=6.5cm]{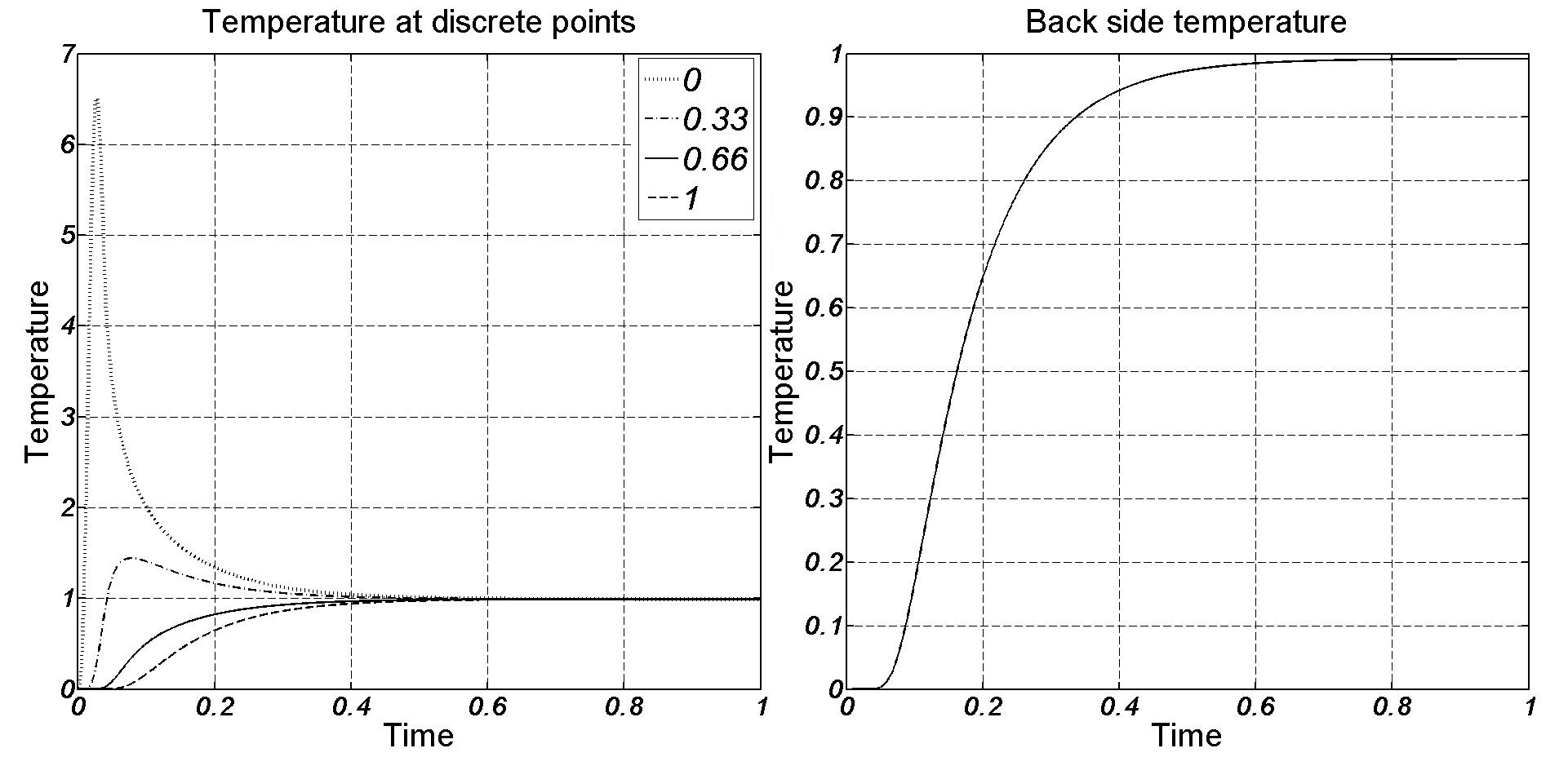}
\caption{Fourier-like solution of the GK-equation, $\tau_\Delta=0.04$, 
$\tau_q=\kappa^2=0.02$}
\label{fig:gkf1}
\end{figure}

Fourier solutions are obtained, if $\tau_q = \kappa^2$. Figure \ref{fig:gkf1} 
shows the case when $\tau_q=\kappa^2=0.02$ and  $\tau_\Delta = 0.04$.

\begin{figure}[ht]
\centering
\includegraphics[width=12cm, height=6.5cm]{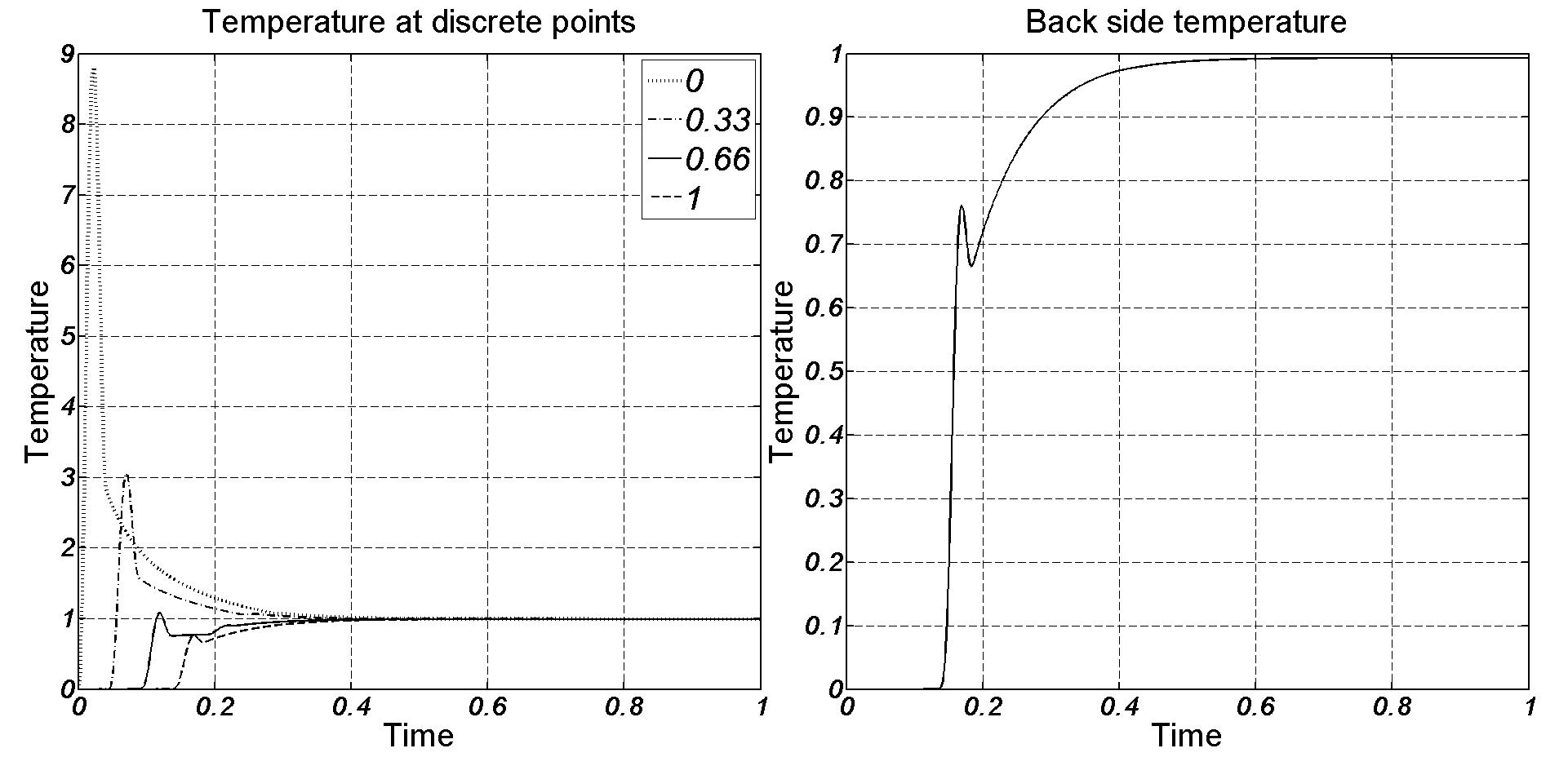}
\caption{MCV-like solution of the GK-equation; $\tau_\Delta=0.04$, 
$\tau_q=0.02$, $\kappa^2=10^{-4}.$}
\label{fig:gkmcv1}
\end{figure}

If the dissipation term is small, one can identify the properties of the 
MCV-equation, see Fig. \ref{fig:gkmcv1}. In this case we 
solved the equation 
for  $\tau_\Delta=0.04$,\ $\tau_q=0.02$ and $\kappa^2=10^{-4}$. $\kappa^2=0$ 
results in exactly the MCV solution. 

\begin{figure}[ht]
\centering
\includegraphics[width=12cm, height=6.5cm]{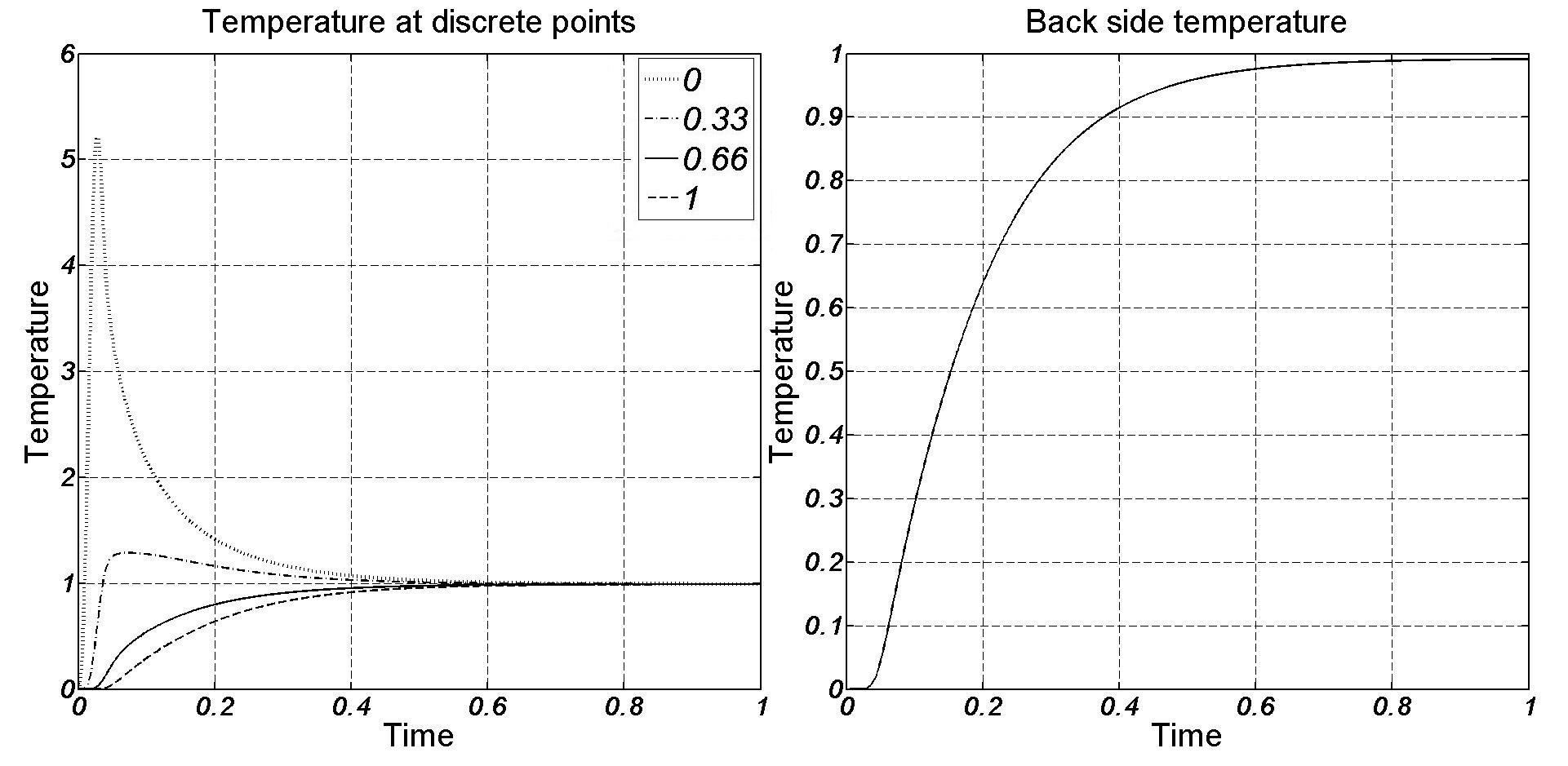}
\caption{Over-diffusive solution of the GK-equation; $\tau_\Delta =0.04,\ 
\tau_q=0.02,\ \kappa^2=0.04.$}
\label{fig:gkgk1}
\end{figure}

In the third case ($\tau_\Delta =0.04,\ \tau_q=0.02,\ \kappa^2=0.04$), the 
solution is over-diffusive. Remarkable is the speed of the signal propagation 
(fig. \ref{fig:gkgk1}). One can see that the temperature signal arrives 
earlier than the measurable signal of the Fourier case and the change of 
the backside temperature is steeper, compared to the Fourier solution. 

\subsection{Solutions of the ballistic-conductive system}
Analysis of the entropy production \re{entrpr} and equation (\ref{genhc}) 
suggests us ideas to distinguish solution classes with respect of 
parameters. From equation (\ref{nd_balcond}) one can see that the parameter 
$\kappa=0$ leads to MCV solutions, because it decouples the last two 
constitutive equations. If it does not hold, there is an additional dissipative 
term in the entropy production. Let us see these cases more closely.
\begin{itemize}
\item \textbf{Fourier solution:}
In this case we need relatively small $\tau_q$ and $\kappa$ but large $\tau_Q$. 
it result in that $\kappa \approx 0$, $\tau_Q+\tau_q \approx \tau_Q$. However, 
we have to note it is only an approximation of Fourier solution, because the 
term of double time derivative of heat flux in equation (\ref{genhc}) never can 
be zero. The particular parameters on Figure \ref{fig:ballfou1} are: 
$\tau_q=0.002; \ \tau_Q=1; \ \tau_{\Delta}=0.04; \ \kappa=0.001$.
\begin{figure}[ht]
\centering
\includegraphics[width=12cm, height=6.5cm]{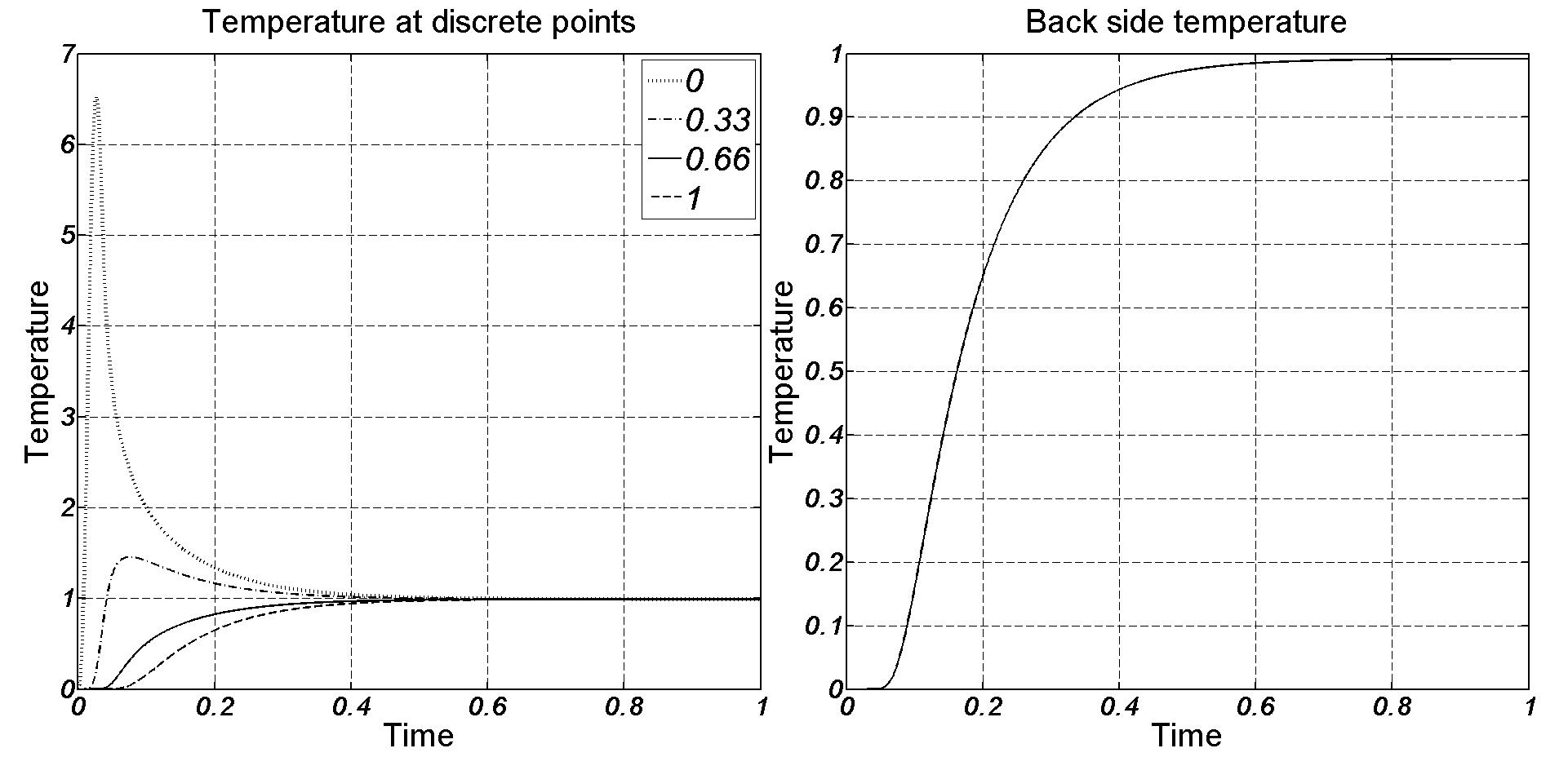}
\caption{Fourier-like solution of the ballistic-conductive system; $\tau_Q=1, \ 
\tau_{\Delta}=0.04,\ \tau_q=0.002,  \ \kappa=0.001$.}
\label{fig:ballfou1}
\end{figure}

\item \textbf{MCV-kind solution:}
This case is very simple, we need only $\kappa=0$ which results in that the 
signal propagation will be independent of $\tau_Q$ and $K=0$. Thus, the 
parameters are: $\tau_{\Delta} = 0.04, \ \tau_q=0.02,\  \kappa=0$. The solution 
is shown on Fig. \ref{fig:ballmcv1}.

\begin{figure}[ht]
\centering
\includegraphics[width=12cm, height=6.5cm]{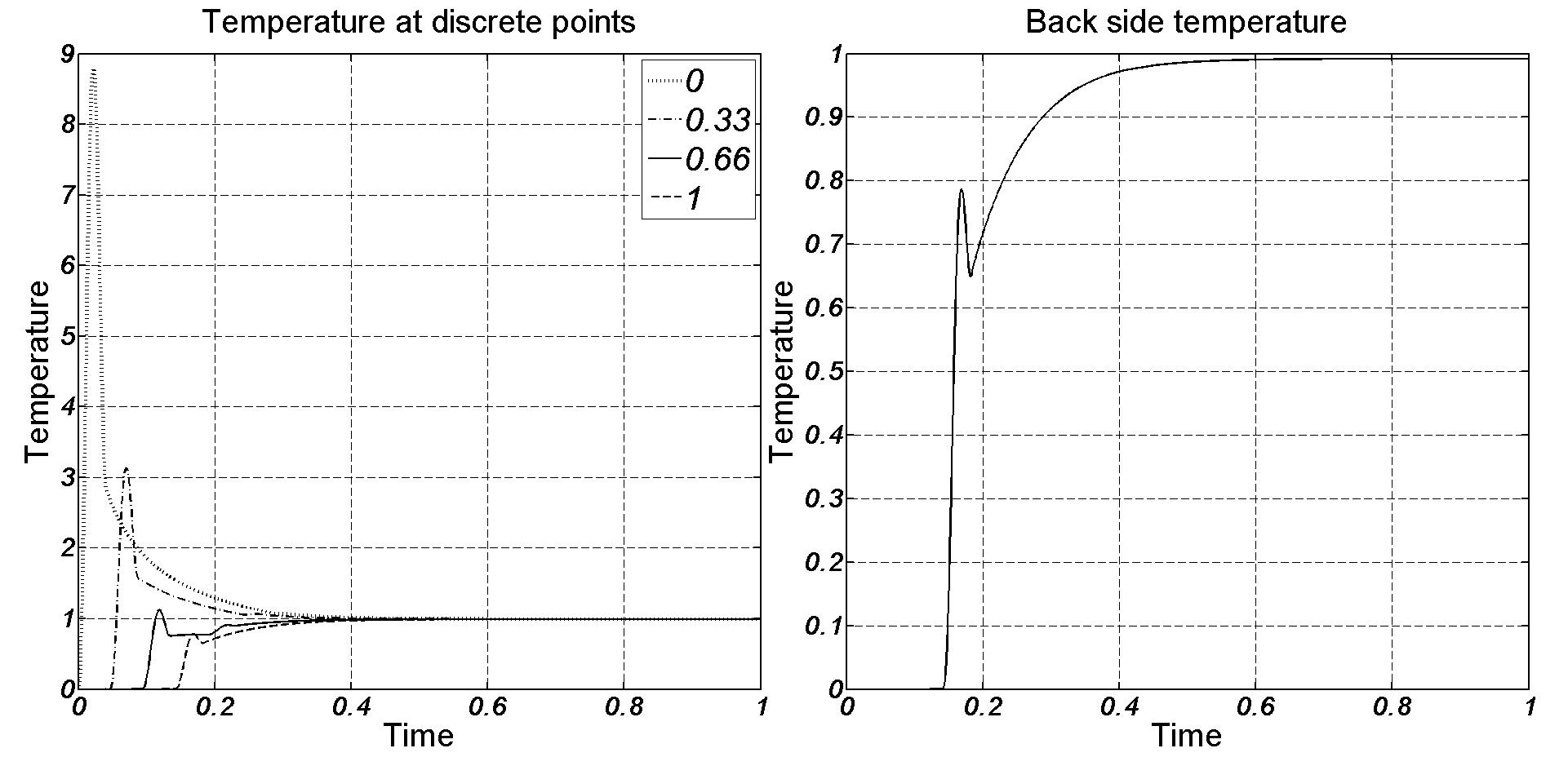}
\caption{MCV-like solution of the ballistic-diffusive system, 
$\tau_{\Delta} = 0.04, \ \tau_q=0.02,\  \kappa=0$}
\label{fig:ballmcv1}
\end{figure}

\item \textbf{GK solution:}
The characteristics are same as before; to reach this we need to apply high 
$\kappa$ and relaxation times are around the same order of magnitude, so we 
used $\tau_Q=0.001,\ \tau_{\Delta}=0.04,\ \tau_q=0.02,\ \kappa=0.25$; see Fig. 
\ref{fig:ballgk1}.

\begin{figure}[ht]
\centering
\includegraphics[width=12cm, height=6.5cm]{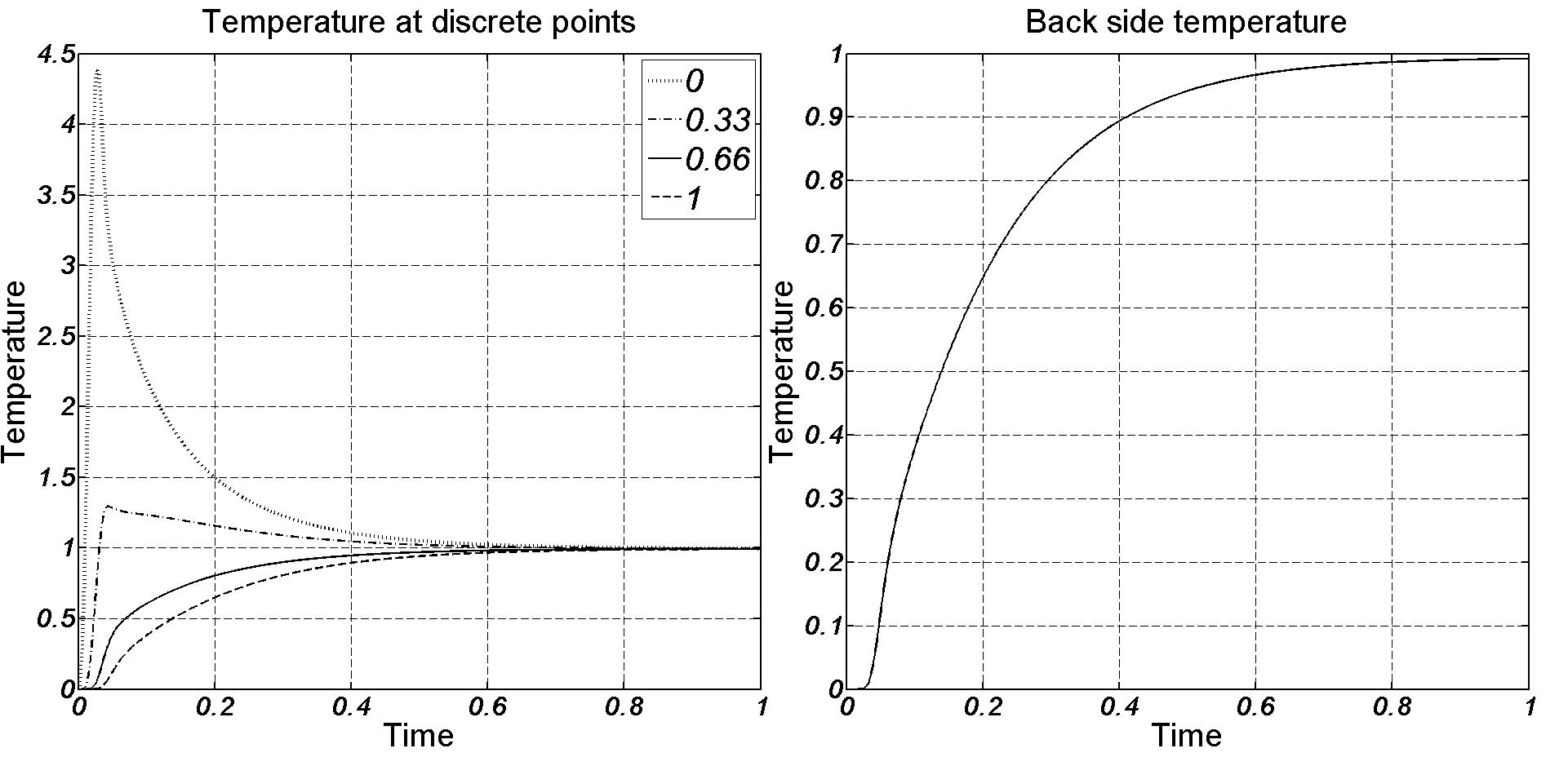}
\caption{Guyer-Krumhansl solution of the ballistic-diffusive system; 
$\tau_Q=0.001,\ \tau_{\Delta}=0.04,\ \tau_q=0.02,\ \kappa=0.25$.}
\label{fig:ballgk1}
\end{figure}

\item \textbf{Ballistic solutions:}
In this case we can observe two propagation speeds, but their detection in  
simulations is not so easy because of dissipation. For example the particular 
parameters $\tau_q=1.9,\ \tau_Q=0.07,\ \kappa=0.28,\ \tau_{\Delta} = 0.065$ 
results in such a propagation which can be seen on Figure 
\ref{fig:ballball1}. 

\begin{figure}[ht]
\centering
\includegraphics[width=12cm, height=6.5cm]{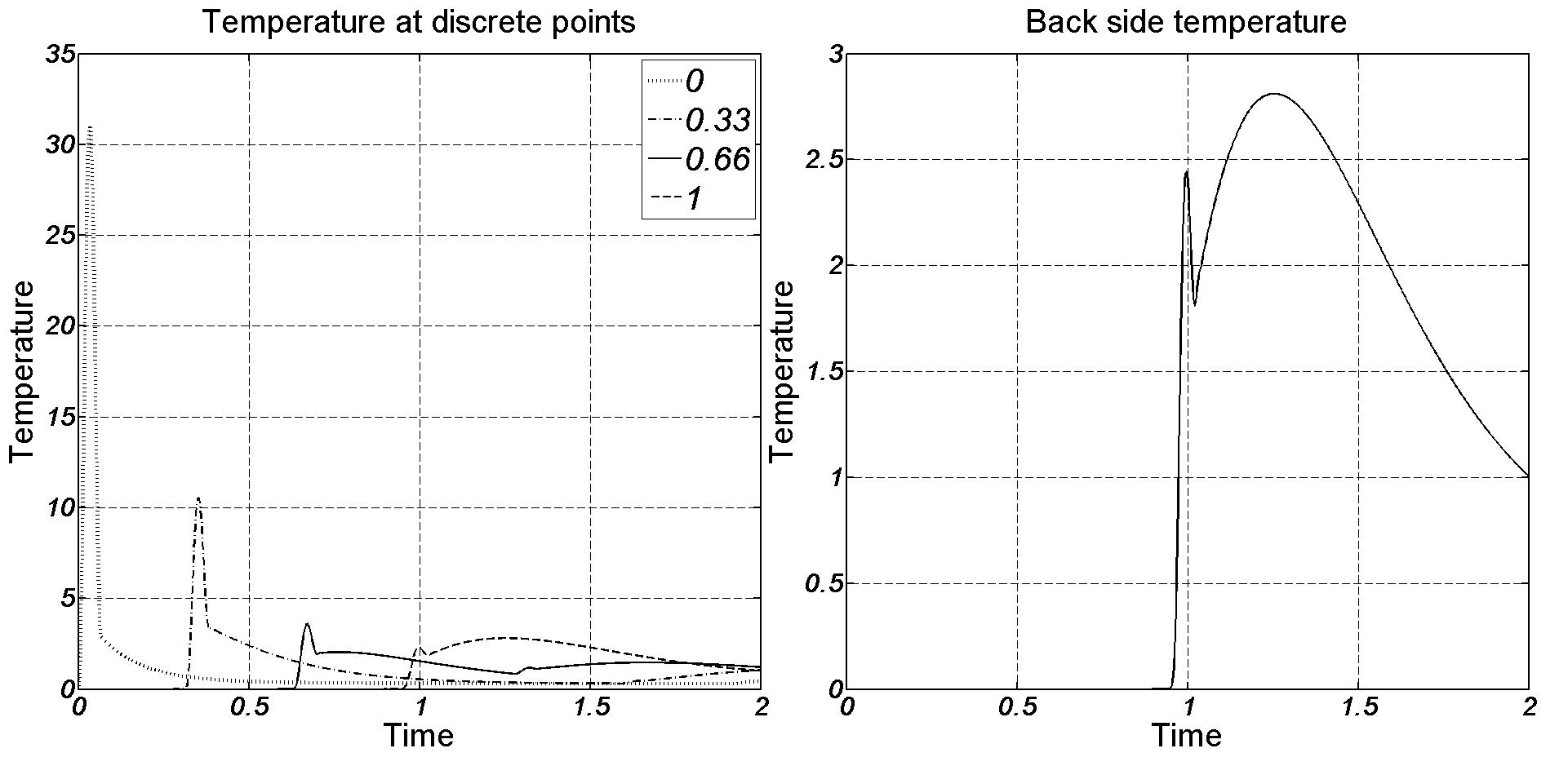}
\caption{Two propagation speeds solving the ballistic-diffusive system;  
$\tau_q=1.9,\ \tau_Q=0.07,\ \kappa=0.28,\ \tau_{\Delta}=0.065$.}
\label{fig:ballball1}
\end{figure}
\end{itemize}

\section{Discussion}

There are some remarkable properties of the thermodynamic approach and the 
obtained heat conduction model.
\begin{itemize}
\item The method of generalization with the help of current multipliers is a 
second order weakly nonlocal extension of the thermodynamic framework and it is 
compatible with the more rigorous exploitation methods of the second law 
\cite{CimVan05a,SelEta13a}. 
\item The thermodynamic thermal force is the gradient of the reciprocal 
temperature and not the gradient of the temperature. Therefore the negative 
temperature solutions of the equations, typical for the MCV and also for 
Jeffreys type equations \cite{Ruk14a}, may be missing.

\item There are several ways to recover Fourier like solutions. Chosing most of 
the material coefficients zero is not the only way. The general heat conduction 
equation results in Fourier solutions in many different coefficient 
combinations, due to the hierarchical structure of the theory. We have 
demonstrated this on the example of the Guyer-Krumhansl equation. We have 
distinguished an under-diffusive or wave like regive, when $\tau_q > \kappa^2$ 
and an over-diffusive regime, when $\tau_q < \kappa^2$. 

These kind of solutions are observed also in case of the Jeffreys type 
equation \cite{TanAra00a}, where it is considered as unphysical 
\cite{SheZha08a}. Let us remark that neither the original dual-phase 
lag model nor related differential constitutive equations are satisfactory from 
a stability point of view \cite{FabLaz14a} and also show other kind of 
inconsistencies \cite{RukSam13a}. Our thermodynamic framework is not 
compatible with the dual-phase lag approach and results in a special case 
results the constitutive equation e \re{genhc5}, that is similar to the 
Jeffreys type one. 

\item The apparent faster than Fourier signal propagation of the over-damped 
case (see fig. \ref{fig:gkgk1}) is a remarkable qualitative effect, contrary to 
the expected phenomena of the MCV equation where the apparent propagation speed 
is slower.

\item Kinetic theory results in correct ballistic propagation using at least 30 
moments, because the material parameters are fixed by the microstructure and 
the introduced particular microscopical processes \cite{DreStr93a}. In our 
phenomenological approach it seems to be no problem to obtain the required 
ballistic wave speeds in the framework of our three field theory, $(e,\bq,\bQ)$.
(13 field in Extended Thermodynamics)
\item The thermodynamic origin of the evolution equations ensures their 
numerical stability and easy solvability.
\end{itemize}

\section{Acknowledgement}

The work was supported by the grants Otka K81161 and K104260. The authors thank 
Bal\'azs Cz\'el, Gyula Gr\'of and Tam\'as F\: ul\: op for valuable 
discussions.

%\bibliographystyle{unsrt}
%\bibliography{termo,VanP,qm,MatolcsiT,nonloc,stmech,misc,fracmech}

\begin{thebibliography}{10}

\bibitem{VanFul12a}
P.~V\'an and T.~F\"ul\"op.
\newblock Universality in heat conduction theory: weakly nonlocal
  thermodynamics.
\newblock {\em Annalen der Physik}, 524(8):470--478, 2012.
\newblock arXiv:1108.5589.

\bibitem{CzeEta13p1}
B.~Cz\'el, T.~F\"ul\"op, Gy. Gr\'of, and P.~V\'an.
\newblock Comparison of temperature responses of the laser flash method in case
  of parabolic and hyperbolic heat conduction models.
\newblock In Dombi Sz., editor, {\em 11th International Conference on Heat
  Engines and Environmental Protection, Balatonfüred}, pages 133--139,
  Budapest, 2013. BME, Dep. of Energy Engineering.

\bibitem{CzeEta13p2}
B.~Cz\'el, T.~F\"ul\"op, Gy. Gr\'of, \'A. Gyenis, and P.~V\'an.
\newblock Simple heat conduction experiments.
\newblock In Dombi Sz., editor, {\em 11th International Conference on Heat
  Engines and Environmental Protection}, pages 141--146, Budapest, 2013. BME,
  Dep. of Energy Engineering.

\bibitem{VanEta13p1}
P.~V\'an, B.~Cz\'el, T.~F\"ul\"op, Gy. Gr\'of, \'A. Gyenis, and J.~Verh\'as.
\newblock Experimental aspects of heat conduction beyond {F}ourier.
\newblock In M.~Pilotelli and G.~P. Beretta, editors, {\em Proceedings of the
  12th Joint European Thermodynamics Conference}, pages 519--524, Brescia,
  2013. Cartolibreria SNOOPY.
\newblock ISBN 978-88-89252-22-2, arXiv:1305.3583.

\bibitem{TiaEta13a}
M.~Tian, N.~Pan, L.~Qu, X.~Guo, and G.~Han.
\newblock A theoretical analysis of local thermal equilibrium in fibrous
  materials.
\newblock {\em Thermal Science}, 2013.
\newblock online first.

\bibitem{SelEta13a}
A.~Sellitto, V.A. Cimmelli, and D.~Jou.
\newblock Entropy flux and anomalous axial heat transport at the nanoscale.
\newblock {\em Physical Review B}, 87(5):054302, 2013.

\bibitem{SelEta15a}
A.~Sellitto, V.A. Cimmelli, and D.~Jou.
\newblock Influence of electron and phonon temperature on the efficiency of
  thermoelectric conversion.
\newblock {\em International Journal of Heat and Mass Transfer}, 80:344--352,
  2015.

\bibitem{Gya77a}
I.~Gyarmati.
\newblock The wave approach of thermodynamics and some problems of non-linear
  theories.
\newblock {\em Journal of Non-Equilibrium Thermodynamics}, 2:233--260, 1977.

\bibitem{Ver97b}
J.~Verh\'as.
\newblock {\em Thermodynamics and {R}heology}.
\newblock Akad\'emiai Kiad\'o and Kluwer Academic Publisher, Budapest, 1997.

\bibitem{Ver83a}
J.~Verh\'as.
\newblock On the entropy current.
\newblock {\em Journal of Non-Equilibrium Thermodynamics}, 8:201--206, 1983.

\bibitem{Nyi91a1}
B.~Ny\'\i{}ri.
\newblock On the entropy current.
\newblock {\em Journal of Non-Equilibrium Thermodynamics}, 16:179--186, 1991.

\bibitem{Van01a2}
P.~V\'an.
\newblock Weakly nonlocal irreversible thermodynamics -- the
  {G}uyer-{K}rumhansl and the {C}ahn-{H}illiard equations.
\newblock {\em Physics Letters A}, 290(1-2):88--92, 2001.
\newblock (cond-mat/0106568).

\bibitem{CimVan05a}
V.~A. Cimmelli and P.~V\'an.
\newblock The effects of nonlocality on the evolution of higher order fluxes in
  non-equilibrium thermodynamics.
\newblock {\em Journal of Mathematical Physics}, 46(11):112901--15, 2005.
\newblock cond-mat/0409254.

\bibitem{Van13p2}
P.~V\'an.
\newblock Thermodynamics of continua: The challenge of universality.
\newblock In M.~Pilotelli and G.~P. Beretta, editors, {\em Proceedings of the
  12th Joint European Thermodynamics Conference}, pages 228--233, Brescia,
  2013. Cartolibreria SNOOPY.
\newblock ISBN 978-88-89252-22-2, arXiv:1305.3582.

\bibitem{AssEta14a}
Cs. Asszonyi, T.~F\"ul\"op, and P.~V\'an.
\newblock Distinguished rheological models for solids in the framework of a
  thermodynamical internal variable theory.
\newblock {\em Continuum Mechanics and Thermodynamics}, 2014.
\newblock accepted, arXiv:1407.0882.

\bibitem{Rug12a}
T.~Ruggeri.
\newblock Can constitutive equations be represented by non-local equations.
\newblock {\em Quarterly of Applied Mathematics}, LXX(3):597--611, 2012.

\bibitem{MulRug98b}
I.~M\"uller and T.~Ruggeri.
\newblock {\em Rational Extended Thermodynamics}, volume~37 of {\em Springer
  Tracts in Natural Philosophy}.
\newblock Springer Verlag, New York-etc., 2nd edition, 1998.

\bibitem{JouAta92b}
D.~Jou, J.~Casas-V\'azquez, and G.~Lebon.
\newblock {\em Extended Irreversible Thermodynamics}.
\newblock Springer Verlag, Berlin-etc., 1992.
\newblock 3rd, revised edition, 2001.

\bibitem{GreNag91a}
A.E. Green and P.~M. Naghdi.
\newblock A re-examination of the basic postulates of thermomechanics.
\newblock {\em Proceedings of the Royal Society: Mathematical and Physical
  Sciences}, 432(1885):171--194, 1991.

\bibitem{BarSte08a}
S.~Bargmann and P.~Steinmann.
\newblock Modeling and simulation of first and second sound in solids.
\newblock {\em International Journal of Solids and Structures}, 45:6067--6073,
  2008.

\bibitem{BarFav14a}
S.~Bargmann and A.~Favata.
\newblock Continuum mechanical modeling of laser-pulsed heating in
  polycrystals: {A} multi-physics problem of coupling diffusion, mechanics, and
  thermal waves.
\newblock {\em ZAMM-Journal of Applied Mathematics and Mechanics/Zeitschrift
  f{\"u}r Angewandte Mathematik und Mechanik}, 94(6):487--498, 2014.

\bibitem{CahHil58a}
J.~W. Cahn and J.~E. Hilliard.
\newblock Free energy of a nonuniform system {I}. {I}nterfacial free energy.
\newblock {\em Journal of Chemical Physics}, 28:258--267, 1958.

\bibitem{ForAme08a}
S.~Forest and M.~Amestoy.
\newblock Hypertemperature in thermoelastic solids.
\newblock {\em Comptes Rendus Mecanique}, 336:347--353, 2008.

\bibitem{KorBer98a}
C.~K\"orner and H.W. Bergmann.
\newblock The physical defects of the hyperbolic heat conduction equation.
\newblock {\em Applied Physics A}, 67:397--401, 1998.

\bibitem{Ruk14a}
Sergey~A Rukolaine.
\newblock Unphysical effects of the dual-phase-lag model of heat conduction.
\newblock {\em International Journal of Heat and Mass Transfer}, 78:58--63,
  2014.

\bibitem{DreStr93a}
W.~Dreyer and H.~Struchtrup.
\newblock Heat pulse experiments revisited.
\newblock {\em Continuum Mechanics and Thermodynamics}, 5:3--50, 1993.

\bibitem{Pes44a}
V.~Peshkov.
\newblock Second sound in {H}elium {II}.
\newblock {\em J. Phys. (Moscow)}, 8:381, 1944.

\bibitem{JacWal71a}
H.~E. Jackson and C.~T. Walker.
\newblock Thermal conductivity, second sound and phonon-phonon interactions in
  {N}a{F}.
\newblock {\em Physical Review B}, 3(4):1428--1439, 1971.

\bibitem{ParEta61a}
W.~J. Parker, R.~J. Jenkins, C.~P. Butler, and G.~L. Abbott.
\newblock Flash method of determining thermal diffusivity, heat capacity, and
  thermal conductivity.
\newblock {\em Journal of Applied Physics}, 32(9):1679, 1961.

\bibitem{BabEta11a}
T.~Baba, N.~Taketoshi, and T.~Yagi.
\newblock Development of ultrafast laser flash methods for measuring
  thermophysical properties of thin films and boundary thermal resistances.
\newblock {\em Japanese Journal of Applied Physics}, 50(11):RA01, 2011.

\bibitem{CzeEta13a}
B.~Cz{\'e}l, K.A. Woodbury, J.~Woolley, and Gy. Gr{\'o}f.
\newblock Analysis of parameter estimation possibilities of the thermal contact
  resistance using the laser flash method with two-layer specimens.
\newblock {\em International Journal of Thermophysics}, 34(10):1993--2008,
  2013.

\bibitem{Jur74b}
Eliahu~Ibrahim Jury.
\newblock {\em Inners and stability of dynamic systems}.
\newblock Wiley New York, 1974.

\bibitem{TanAra00a}
DW~Tang and N~Araki.
\newblock Non-fourier heat condution behavior in finite mediums under pulse
  surface heating.
\newblock {\em Materials Science and Engineering: A}, 292(2):173--178, 2000.

\bibitem{SheZha08a}
B~Shen and P~Zhang.
\newblock Notable physical anomalies manifested in non-{F}ourier heat
  conduction under the dual-phase-lag model.
\newblock {\em International Journal of Heat and Mass Transfer},
  51(7):1713--1727, 2008.

\bibitem{RukSam13a}
S.~A. Rukolaine and A.~M. Samsonov.
\newblock Local immobilization of particles in mass transfer described by a
  {J}effreys-type equation.
\newblock {\em Physical Review E}, 88:062116, 2013.

\bibitem{FabLaz14a}
M.~Fabrizio and B.~Lazzari.
\newblock Stability and second law of thermodynamics in dual-phase-lag heat
  conduction.
\newblock {\em International Journal of Heat and Mass Transfer}, 74:484--489,
  2014.

\end{thebibliography}

\end{document}